\documentclass[aps,floatfix,prd,preprint]{revtex4}
\usepackage{graphicx} 
\usepackage{bm}
\usepackage{amsmath}
\usepackage{float}
\usepackage{longtable}
\usepackage{braket}
\usepackage{comment}
\usepackage{longtable,array,booktabs}

\def\be{\begin{equation}}
\def\ee{\end{equation}}
\def\A{{\bf A}}
\def\B{{\bf B}}
\def\x{{\bf x}}

\def\y{{\bf y}}

\def\D{{\bf D}}

\newcommand{\bnabla} {{\mbox{\boldmath $\nabla$}}}

\begin{document}

\title{Strong Decays of the Light Exotic $0^{+-}$ and $2^{+-}$ Hybrid Mesons}
\author{Christian Farina}\email[]{cfarina@illinois.edu}
\affiliation{Department of Physics, University of Illinois, Urbana-Champaign, IL 61801, USA.}
\author{Eric S.\,Swanson}\email[]{swansone@pitt.edu}
\affiliation{Department of Physics and Astronomy, University of Pittsburgh, Pittsburgh, PA 15260, USA.}

\begin{abstract}
A model of hybrid meson structure based on the QCD Hamiltonian in Coulomb gauge and  a single constituent quasigluon is applied to the computation of hadronic decays of mesons with exotic quantum numbers, $0^{+-}$ and $2^{+-}$.  These correspond to hybrid mesons in which the gluon couples to a $q\bar{q}$ pair in an $P$-wave and can therefore be identified as orbital excitations of the exotic $1^{-+}$ state. Interestingly, we find $0^{+-}$ states to be narrow, contrary to what was found by previous calculations. This is primarily due to the suppression of the decay mode $0^{+-}\rightarrow 2^1S_0+1^1S_0$, which is unique to our model. The $2^{+-}$ states are found to be narrow, as expected.
\end{abstract}

\maketitle

\section{Introduction}

Hybrid mesons have generated interest for decades  because they can provide a portal to nonperturbative gluodynamics. Exotic hybrids -- those whose quantum numbers cannot be realized by a $q\bar{q}$ pair -- are particularly interesting because they do not mix with conventional $q\bar{q}$ states and their  detection would be a smoking-gun signature of new physics.

Indeed, several experimental efforts such as the GlueX at JLab, PANDA at FAIR and BESIII at BEPC II will be, or are currently, under way and there is hope that  future  data, combined with continuous progress in lattice field computations, will shed light on these enigmatic states. However, the field is not without progress. The puzzle of the two exotic $1^{-+}$ states labeled $\pi_{1}(1400)$ and $\pi_{1}(1600)$ \cite{Workman:2022ynf} seems to have mainly been resolved, suggesting that only the heavier of the two is necessary to explain the data \cite{JPAC:2018zyd,Kopf:2020yoa}. Recently, the BESIII collaboration \cite{BESIII:2022riz} has observed an isoscalar resonance with exotic quantum numbers $1^{-+}$ in $J/\psi \to \gamma \eta \eta'$. This new state, called $\eta_1(1855)$, has a mass of 
$1855 \pm 9 {}^{+6}_{-1}$ MeV and a width of $188 \pm 18 {}^{+3}_{-8}$  MeV, making it an ideal candidate for the isoscalar partner to the isovector $\pi_1(1600)$ \cite{BESIII:2022riz, Swanson:2023zlm}.

Hybrid mesons models have a long history, dating back to the 1970s, and can be broadly divided into two categories: flux-tube models and constituent gluon models (see \cite{Meyer:2015eta} for a review). These two camps primarily  differ in how they describe the gluonic degrees of freedom. In flux-tube models \cite{flux} (which include string models \cite{string}), glue is represented as a collective string-like excitation, while in constituent models the gluonic degrees of freedom are described by  quasiparticles. The latter notion finds confimation in recent lattice studies demonstrating that the low-lying charmonium hybrid spectrum is well-described by a single gluonic quasiparticle with axial quantum numbers $J_G^{PC}=1^{+-}$ \cite{Dudek:2011bn}.

Throughout the years, constituent gluon models have developed from almost dynamics-free models \cite{HM} to more modern pictures which include many of the features of QCD \cite{Swanson:1997wy,Swanson:1998kx,Guo:2008yz}. More recently, the constituent picture with the QCD Hamiltonian in Coulomb gauge as a starting point has been applied to charmonium hybrid decays \cite{Farina:2020slb}, the light $H_{1}$ states ($1^{--}$, $(0,1,2)^{-+}$) \cite{Farina:2023oqk}, and to a study of light hybrid meson mixing \cite{Swanson:2023zlm}. In this paper, we use this picture to compute the strong decays of the light exotic $0^{+-}$, $2^{+-}$ and $2^{+-'}$ hybrid mesons. These correspond to hybrid mesons in which the gluon couples to a $q\bar{q}$ pair in an $P$-wave and can therefore be identified as the orbital excitations of the exotic $1^{-+}$ state previously studied in \cite{Farina:2023oqk}.

The paper is organized as follows: Section \ref{sec:QCD} contains  details of the QCD Hamiltonian in Coulomb gauge, which forms the basis of our model, and the construction of the light hybrid meson state from constituent quasiparticles. The model parameters used in the computation and the decay model are also contained in this section, along with the resulting selection rules. Section \ref{sect:dec} presents the results obtained for both isoscalar and isovector hybrids. Summary and conclusions, along with prospects for potential discovery of these states, are found in Section \ref{sect:conc}.

\section{Constituent-Gluon Model of Hybrids}
\label{sec:QCD}
\subsection{QCD Hamiltonian In Coulomb Gauge}

The starting point of our model is the QCD Hamiltonian in Coulomb gauge. The main advantages of this gauge choice are that all degrees of freedom are manifestly physical and that an instantaneous ``Coulomb" interaction between constituent quasiparticles emerges naturally. This choice is justified by previous work by Swanson and Szczepaniak \cite{Szczepaniak:2001rg}, who, starting from the QCD Hamiltonian in Coulomb gauge, developed a mean-field model of the gluonic vacuum and obtained an effective quasigluon dispersion relation and an approximation for the vacuum expectation value of the Coulomb kernel $K^{AB}$ described below. For those interested in the details of the QCD Hamiltonian in Coulomb gauge and its derivation, the standard reference is \cite{Christ:1980ku}. Here we just highlight the main features of the model. The full Hamiltonian is given by 
\begin{eqnarray}
H&=&\,\int\,\left[\psi^{\dagger}\left( -i\bm{\alpha}\cdot\bm{\nabla}+\beta m\right)\psi+\frac{1}{2}\left(\mathcal{J}^{-1/2}\,\mathbf{\Pi}^{A}\,\mathcal{J}\,\cdot \mathbf{\Pi}^{A}\,\mathcal{J}^{-1/2}\,+\mathbf{B}^{A}\cdot \mathbf{B}^{A}\right)\right]\,d^{3}x\, \nonumber \\
&-&g\int \psi^{\dagger}\bm{\alpha}\cdot\mathbf{A}\psi  d^{3}x+H_C,
\label{eqn:QCDHCoulomb}
\end{eqnarray}
with
\begin{equation}
H_C = \frac{1}{2}\int {\cal J}^{-1/2} \rho^A(\x)
 {\cal J}^{1/2} \hat K^{AB}(\x,\y;\A) {\cal J}^{1/2} \rho^B(\y) {\cal J}^{-1/2}d^3x\, d^3y.
\label{eq:hc}
\end{equation}
In the above, ${\cal J}$ is the Faddeev-Popov determinant, written as ${\cal J} \equiv {\rm det}(\bnabla\cdot \D)$, and $\D$ is the adjoint covariant derivative, 
 $\D^{AB} \equiv \delta^{AB} \bnabla  - g f^{ABC}\A^C$.
The color charge density contains both quark and gluon components
\begin{equation}
\rho^A({\bf x}) =
 f^{ABC} {\bf A}^B({\bf x}) \cdot {\bf \Pi}^C({\bf x}) + \psi^{\dag}(\x)T^A\psi(\x),
\label{eq:rho}
\end{equation}
while the kernel of the Coulomb interaction can be formally written as
\begin{equation}
\hat K^{AB}({\bf x},{\bf y};\A) \equiv \langle{\bf x},A|
 \frac{ g }{ \bnabla\cdot {\bf D}}(-\bnabla^2)
 \frac{ g }{ \bnabla\cdot {\bf D}}|{\bf y},B\rangle.
\label{eq:K}
\end{equation}
The Coulomb kernel is instantaneous and contains all powers of gluonic field operators. These may be organized in a tower of Dyson equations, and truncations can be solved. This approach shows that confinement appears with the aid of a simple variational vacuum ansatz\cite{Szczepaniak:2001rg}. However, this potential does not necessarily need to be equivalent to the interaction between static color sources, which is typically used in phenomenological calculations\cite{Dawid:2024uqz}. We therefore choose to simplify by replacing the vacuum matrix element of the Coulomb interaction with:
\begin{equation}
  K^{AB}(\mathbf{r},0)=\delta^{AB} \left( \frac{a_s}{r}-\frac{3}{4}br+\mathcal{C}\right), 
  \label{eq:K}
\end{equation}
where $a_s$, $b$, and $\mathcal{C}$ are parameters to be determined by fitting to the meson spectrum, as explained below. This choice, besides reproducing the well-known Cornell potential of non-relativistic meson models, finds support in lattice simulations of the $q\bar{q}$ static quark potentials with the gluonic field in different configurations, which show that they all exhibit a long-range linear behavior (see, for example, Ref.  \cite{Juge:1997nc}).

In the above, the gluon field is approximated by the perturbative expression:
\begin{equation}
    \mathbf{A}^{B}(\mathbf{x})=\int \frac{d^3k}{(2\pi)^3}\frac{1}{\sqrt{2\omega(k)}}\left(a^{B}(\mathbf{k})+a^{B}(\mathbf{-k}) \right) e^{i\mathbf{k \cdot x}}.
\end{equation}
The quasigluon dispersion relation differs from the perturbative $\omega=k$ and the mean-field calculation discussed above finds that it can be well-approximated by the ansatz
\begin{equation}
    \omega^2(k)=k^2+m_{g}^{2}e^{-k/b_g},
    \label{eq:omega}
\end{equation}
where $m_g$ is the dynamical gluon mass given by $m_g\approx 600$ MeV and $b_g$  is a parameter with value $b_g \approx 6000$ MeV. We remark that despite the appearance of an effective mass, the quasigluon remains transverse.

\subsection{Light Hybrids}

Our model of hybrid structure assumes that the dominant degrees of freedom are constituent quarks and dynamical constituent gluons with interactions specified by Eqs. ref{eqn:QCDHCoulomb} and ref{eq:hc}. The hybrid meson state is constructed, by assumption, from the  vacuum Fock state ($\ket{0}$) as a bound state of the minimal number of constituent quasiparticles: a quark-antiquark pair coupled with a valence axial gluon in a specific color, spin, and orbital momentum configuration. This assumption is made plausible here by the relatively large gluon mass. We work in momentum space and employ Jacobi coordinates. 

The total meson angular momentum $J$ is obtained by first coupling the gluon spin projection to the gluon angular momentum, $\ell_g$, to obtain the total gluon angular momentum $j_g$. Calculations can be streamlined by working in the gluon helicity basis, which is achieved by means of the relationship

\begin{equation}
a^{\dagger, A}_{m_g}(\bm{k}) = \mathcal{D}^{(1)*}_{m_g,\lambda}(\hat{k})a^{\dagger, A}_{\lambda}(\bm{k}),
\end{equation}
where $a^{\dagger}$ is a constituent gluon creation operator, $m_g$ is the $z$-component of the gluon angular momentum in the canonical basis and $\lambda$ is the helicity of the gluon. $\mathcal{D}^{(1)*}_{m_g,\lambda}$ is the Wigner-$\mathcal{D}$ matrix which rotates the operator from the canonical basis to the helicity basis. Doing so and assuming  that  $\ell_g = j_g$ produces a Clebsch-Gordan factor of
\be
\chi^{(-)}_{\lambda,\mu} \equiv \langle 1 \lambda \ell_g 0| \ell_g \mu\rangle =
\begin{cases} 0, \ell_g = 0 \\ \frac{\lambda}{\sqrt{2}} \delta_{\lambda,\mu}, \ell_g \geq 1 \end{cases}.
\ee
This represents a transverse electric (TE) gluon and forms the explicit realization of the axial constituent gluon. Setting $\ell_g = j_g\pm 1$ yields a transverse magnetic (TM) gluon with a Clebsch factor given by $\chi^{(+)}_{\lambda,\mu} = \delta_{\lambda,\mu}/\sqrt{2}$. Combining with quark spins yields the final generic expression for a hybrid ket
\begin{align}
&|JM [LS \ell j_g \xi]\rangle = \frac{1}{2} T_{ij}^A\int \frac{d^3q}{(2\pi)^3}\, \frac{d^3k}{(2\pi)^3} \,
\Psi_{j_g;\ell m_\ell}({\bf k}, {\bf q})\, \sqrt{\frac{2 j_g +1}{4\pi}} \, D_{m_g\mu}^{j_g*}(\hat k) \, \chi^{(\xi)}_{\mu,\lambda} \nonumber \\
& \times
\langle \frac{1}{2} m \frac{1}{2} \bar{m} | S M_{S} \rangle \,
\langle \ell m_\ell, j_g m_g| L M_L\rangle \,
\langle S M_S, L M_{L} | J M \rangle \,
b_{{\bf q}-\frac{{\bf k}}{2},i,m}^\dagger \,
  d_{-{\bf q}-\frac{{\bf k}}{2},j, \bar{m}}^\dagger \,
a^\dagger_{{\bf k}, A, \lambda} |0\rangle.
\label{eq:PSI}
\end{align}
The $T^{A}_{ij}$ in the above equation is the color factor, which is obtained by assuming the $q\bar{q}$ to be in a color-octet coupled to the gluon so as to give an overall color singlet, as demanded by color confinement. The explicit form of $\Psi_{j_g;\ell m_\ell}({\bf k,\bf q})$ is given in Eq. \ref{eq:radWavefunct} below. Since all lowest lying hybrids have $\xi=-1$ and $j_g=1$, only the TE mode is of interest for this work.
Furthermore, the  hybrid state is an eigenstate of parity and charge conjugation with eigenvalues given by
\be
P =\xi (-1)^{\ell + j_g+1}  \ \ \textrm{and} \  \ C= (-1)^{\ell+S+1}.
\ee

\subsection{Model Parameters}
\label{sect:model}

Many parameters go into the computation of the hybrid decay amplitudes: the interaction parameters of Eq. \ref{eq:K}, the quark masses, and the gluon parameters. The spin-dependent interaction used to determine quark parameters, which was obtained from Ref.~\cite{Swanson:2023zlm}, is reproduced below for completeness:
\begin{align}
V_{SD} &= 2C_F \frac{\alpha_H}{3 r m_q m_{\bar q}}\,  S_q \cdot S_{\bar q} \, b_0^2\exp(-b_0 r) \, \theta(r_0)\nonumber \\
       &+ \frac{1}{2}\left(C_F \frac{\alpha_H}{r^3}  + (2\epsilon-1)\frac{b}{r}\right) \cdot \left(\frac{S_q\cdot L}{m_q^2} + \frac{S_{\bar q}\cdot L}{m_{\bar q}^2}\right)\, \theta(r_0) + \left(C_F\frac{\alpha_H}{r^3} + \epsilon\frac{b}{r}\right) \frac{S\cdot L}{m_q m_{\bar q}}\, \theta(r_0) + \nonumber \\
       &+ \frac{4\alpha_H}{3 m_q m_{\bar q} r^3} \left( S_q\cdot r \, S{\bar q}\cdot r - S_q \cdot S_{\bar q}\right) \, \theta(r_0).
       \label{eq:SDmodel}
\end{align}
Details about the various terms in this interaction can be found in Ref. \cite{Swanson:2023zlm}. All parameters are obtained by fitting light \emph{uds} meson masses and the values obtained from the fit, which are used in this work, are reported in Table \ref{tab:models2}.

\begin{table}[h]
\caption{Model Parameters. The values were obtained by fitting the entire isovector and isoscalar \emph{uds} spectrum. The values of $m_{n}$ and $m_{s}$ correspond to the light and strange quark masses, respectively.}
\scalebox{0.8}{
\begin{tabular}{cccccccc|cc}
\hline\hline
 $m_n\ (m_s)$ (MeV) & $a_S$ & $\sigma$ (GeV$^2$) & $\mathcal{C}$ (MeV) & $\alpha_H$ & $b_0$ (GeV$^{-1}$) & $r_0$ (GeV$^{-1}$) & $\epsilon$ (GeV$^{-1}$) & rel. error & avg deviation (MeV)\\
\hline
420 (606) & 1.496 & 0.053 & 0.149 & 1.999 & 0.389 & 9.1 & 0.25 & 6\% & 69 \\
\hline\hline
\end{tabular}}
\label{tab:models2}
\end{table}

Once parameters are fixed, the light hybrid spectrum can be computed. The hybrid bound state equation is obtained from the  QCD Hamiltonian by computing the expectation value, 
\begin{equation}
    \langle J'M'[L'S'\ell' j' \xi']| H | JM[LS \ell j \xi]\rangle.
    \label{eq:expValue}
\end{equation}
Given that our immediate goal is to obtain spin-independent multiplets, we take the nonrelativistic limit of the currents in Eq. \ref{eq:hc}.  The resulting spectrum can be categorized according to interpolating operators, as indicated in Table \ref{tab:JPC} \cite{Juge:1997nc}. Here $\B$ is the chromomagnetic field, and $\psi$ and $\chi$ are heavy quark and antiquark fields, respectively. The remaining columns give the corresponding quantum numbers and the hybrid meson $J^{PC}$ quantum numbers in the specified multiplet. 
\begin{table}[h]
\caption{$J^{PC}$ Hybrid Multiplets.}
\begin{tabular}{c|c|cccc|l}
\hline\hline
multiplet & operator & $\xi$ & $j_g$ & $\ell$ & $L$ & $J^{PC}\ S=0\ (S=1)$ \\
\hline
$H_1$ & $\psi^\dagger \B \chi$ & -1 & 1 & 0 & 1 & $1^{--}$, $(0,1,2)^{-+}$ \\
$H_2$ & $\psi^\dagger \bnabla \times \B \chi$  & -1 & 1 & 1 & 1 & $1^{++}$, $(0,1,2)^{+-}$ \\
$H_3$ &$\psi^\dagger \bnabla \cdot \B \chi$ & -1 & 1 & 1 & 0 & $0^{++}$, $(1^{+-})$ \\
$H_4$ & $\psi^\dagger [\bnabla \B]_2 \chi$ & -1 & 1 & 1 & 2 & $2^{++}$, $(1,2,3)^{+-}$ \\
\hline\hline
\end{tabular}
\label{tab:JPC}
\end{table}

The gluon parameters used were the ones reported in the gluon dispersion relationship of Eq. \ref{eq:omega} ($m_g=600$ MeV and $b_{g}=6000$ MeV). 

The Hamiltonian expectation value of Eq. \ref{eq:expValue} yields integral expressions in which the angular parts can be done analytically, leaving only radial integrals which can be computed in either momentum or configuration space (for details see \cite{Farina:2020slb}). The only matter left to settle is then the choice of basis in which to diagonalize resulting radial system. There are many possible choices and all come with their own advantages and pitfalls. We choose to expand the radial wavefunctions in a finite Gaussian basis of the form $
\Psi_{j_g,\ell}(\vec k,\vec q) \propto \Sigma_{n,m}D^{j_g*}_{m_g,\mu}(\hat k) k \exp(-k^2n^2/2\beta^2) \cdot Y_{\ell m_\ell}(\hat q) q^\ell \exp(-q^2m^2/2\alpha^2)$. This choice is in line with what has been historically done for conventional meson and baryon models, in which authors have used empirical Gaussian expansion methods as an alternative to SHO and hyperspherical methods \cite{Richard:1992uk}. 

The main advantage of this basis is that it yields analytical results for all the integrals in the Hamiltonian expectation value above. Additionally, the expression can be easily Fourier-transformed, allowing the computation in either the configuration or the momentum basis.

Here, however, since the main goal is to compute the decay properties of the mesons,  we are only interested in the gross features of meson structure and the resulting hybrid wavefunctions are therefore approximated by a product of two Gaussians  (this sets $n=m=1$ in the expansion above) with widths $\alpha$ and $\beta$:
\be
\Psi_{j_g,\ell}(\vec k,\vec q) \propto D^{j_g*}_{m_g,\mu}(\hat k) k \exp(-k^2/2\beta^2) \cdot Y_{\ell m_\ell}(\hat q) q^\ell \exp(-q^2/2\alpha^2).
\label{eq:radWavefunct}
\ee
The parameters $\alpha$ and $\beta$ are then obtained by minimizing the expectation value of Eq. \ref{eq:expValue}:
\begin{equation}
  \frac{\partial}{\partial\alpha(\beta)} \braket{J'M'[L'S'\ell' j' \xi']| H | JM[LS \ell j \xi]}=0
\end{equation}
Similarly, ordinary mesons are described by SHO wavefunctions depending on a scale parameter $\beta_m$. This parameter is chosen to give the best overlap with the full numerically determined ordinary meson wavefunctions.

\subsection{Decay Model}

\begin{figure}
    \centering
    \includegraphics[width=0.8\linewidth]{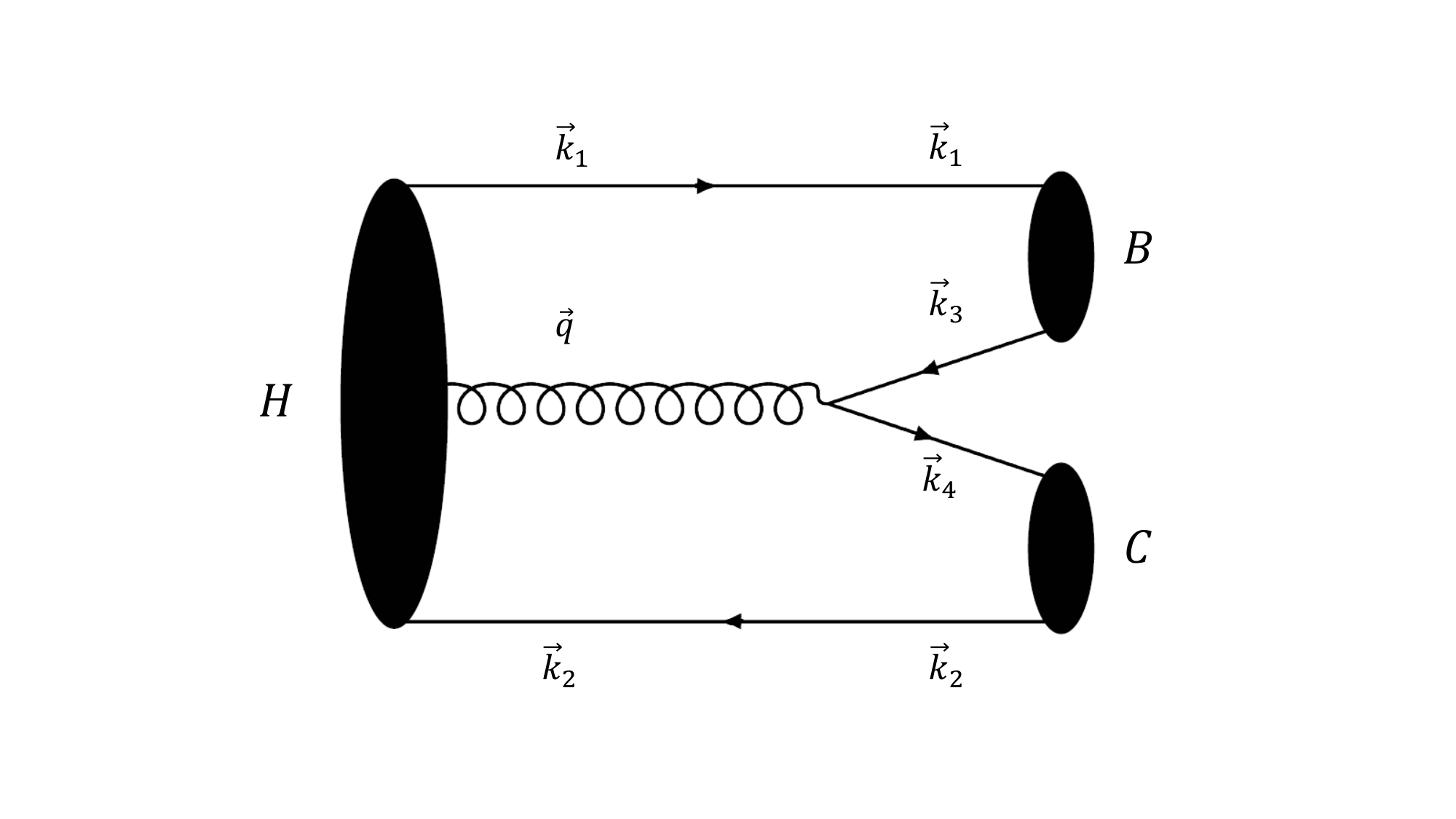}
    \caption{Schematic diagram of the hybrid decay model, in which the hybrid meson decays into two conventional mesons through gluon dissociation into a $q\bar{q}$ pair.}
    \label{fig:hdecay}
\end{figure}

With a model of hybrid structure at hand, one can develop a model of hybrid decays as well. The idea is to assume that the decay takes place through the simplest leading gluon operator, namely quark-pair production through gluon dissociation (see Fig. \ref{fig:hdecay}).

This  decay model yields two selection rules that are typical of hybrid decay models, namely: (i) spin zero hybrids do not decay to spin zero mesons, (ii) TE
hybrids do not decay to mesons with identical spatial wavefunctions. The former follows from the $^3S_1$ quantum
numbers of the vertex, while the latter follows from the symmetry properties of the decay amplitude (see Ref. \cite{Farina:2020slb} for details). Additionally, decays to an $S$-wave and $P$-wave meson pair are generally dominant.

\section{Light Hybrid Decays}
\label{sect:dec}

\subsection{Hybrid Masses}

Hybrid masses constitute another source of uncertainty in the model. Because our model only gives spin-averaged hybrid masses, and there are currently no experimental data to constrain them, one must rely on lattice data. However, the results from lattice gauge theory are somewhat scattered. For example, a study of the lightest hybrids supermultiplet \cite{Dudek:2011bn} computes isovector masses of all the $H_1$ hybrids and the exotic $0^{+-}$ and $2^{+-}$ hybrids at different pion masses, ranging from $m_{\pi}\approx700$ MeV down to $m_{\pi}\approx400$ MeV. A crude extrapolation down to the physical pion mass gives estimates for the isovector hybrids but not the isoscalars.
Alternatively, a more comprehensive study of the full isovector-isoscalar meson spectrum \cite{Dudek:2013yja} extracted hybrid masses at $m_{\pi}\approx400$ MeV. The results suggest that there should be a $1^{-+}$ isovector at about $2025$ MeV, and two isoscalar counterparts at higher masses of about $2150$ MeV and $2340$ MeV, respectively. The excited state masses are obtained in the same way, and they show an average mass splitting of about 35 MeV
between the isovector and low isoscalar and about 100 MeV between the low and high isoscalar.
However, the light isovector $1^{-+}$ mass of about 2025 MeV, is roughly $400$ MeV higher than the accepted $\pi_1(1600)$ mass. The approach adopted here, admittedly not overly rigorous, is to shift all of the masses in Ref. \cite{Dudek:2013yja} down by the same amount (425 MeV) so that the isovector $1^{-+}$ mass matches the $\pi_1(1600)$ mass. This keeps the mass splittings the same and gives the masses for the excited states reported in the Table \ref{tab:hybridMasses}. 
\begin{table}[ht]
    \caption{Exotic hybrid masses reported in MeV. The numbers on the left in each column are the lattice values taken from \cite{Dudek:2013yja}. Numbers in bold are the shifted masses.}
    \begin{tabular}{c|c|c|c}
    \hline\hline
     $J^{PC}$ & ~~~Isovector~~~ & ~Low Isoscalar~ & ~High Isoscalar~ \\
     \hline
     $1^{-+}$ & 2025 (\textbf{1600}) & 2155 (\textbf{1730}) & 2340 (\textbf{1915}) \\
     $0^{+-}$ & 2325 (\textbf{1900}) & 2380 (\textbf{1955}) & 2480 (\textbf{2055}) \\
     $2^{+-}$ & 2440 (\textbf{2015}) & 2460 (\textbf{2035}) & 2550 (\textbf{2125}) \\
     $2^{+-'}$ & 2650 (\textbf{2225}) & 2680 (\textbf{2255}) & 2760 (\textbf{2345}) \\
     \hline\hline
    \end{tabular}
    \label{tab:hybridMasses}
\end{table}

\subsection{Isovectors}
\label{sect:isovectors}
In order to compute hybrid decays into kaons, one must first determine the mixing angle for the spin-one kaons, which are defined as linear superpositions of spin singlet and triplet states through a mixing angle $\theta_{K}$:
\begin{eqnarray}
|K_{1L}\rangle&=&\cos{\theta_{K}}|^1P_1\rangle+\sin{\theta_{K}}|^3P_1\rangle \nonumber \\
|K_{1H}\rangle &=&-\sin{\theta_{K}}|^1P_1\rangle+\cos{\theta_{K}}|^3P_1\rangle.
\end{eqnarray}
There is ongoing theoretical debate about the value of the mixing angle \cite{Suzuki:1993yc, Cheng:2011pb, Hatanaka:2008gu}, but no experimental data to definitively constrain it. 
The authors of \cite{Cheng:2011pb} argue that the value of $|\theta_{K}|$ can be obtained by first determining the  mixing angles for $f_{1}(1285)/f_{1}(1450)$ and $h_{1}(1170)/h_{1}(1380)$ from  mass relations. These angles depend on the masses of the $K_{1L}$ and $K_{1H}$, which in turn depend on the mixing angle $\theta_{K}$. This procedure gives a value of $\theta_{K}=35^\circ$. Alternatively, Suzuki determined the mixing angle from  partial decay rates and, independently, from  masses, finding that the angle could be either $\theta_{K}=33^\circ$ or $\theta_{K}=57^\circ$, but that $\theta_{K}=33^\circ$ is favored by the observed production dominance of $K_{1}(1400)$ in  $\tau$ decays\cite{Suzuki:1993yc}. This is the value we employ in the following.

Lastly, mixing angles for the $\eta$ and $\eta'$ are specified by the simple ansatz:
\begin{eqnarray}
|\eta\rangle &=&\frac{1}{2}(|u\bar u\rangle + | d \bar d\rangle )-\frac{1}{\sqrt{2}}|s\bar{s}\rangle \nonumber \\
|\eta'\rangle &=&\frac{1}{2}(|u\bar{u}\rangle + | d \bar{d}\rangle )+\frac{1}{\sqrt{2}}|s\bar{s}\rangle.
\end{eqnarray}

This completes the specification of the decay model. Resulting predictions for the decay of the isovector $0^{+-}$,  $2^{+-}$, and $2^{{+-}'}$  hybrids  are presented in 
 Table \ref{tab:IsovectorDecays}.

\begin{longtable}{c|c|c|ccc}
\caption{Strong decays of the three isovector states. The number in parenthesis next to the hybrid multiplet is the hybrid mass. The meaning of the symbols is as follows: x = negligible, $\o=$ threshold, -- = forbidden.}
\label{tab:IsovectorDecays}\\
\hline\hline
~~Hybrid $J^{PC}$~~& ~Decay Channel~ & ~Wave~ & ~~$H_2~(1.900~$GeV)~~  &   \\
\hline
$~0^{+-}~$ & $h_1(1170)\pi$ & $P$ & 56.6 &  & \\
         & $\pi(1300)\pi$ & $S$ & 2.5 &  \\
         & $\omega(1420)\pi$ & $P$ & x &  \\
         & $b_1(1235)\eta$ & $P$ &  17.6 &  & \\
         & $K_{1L}\bar{K}$ & $P$ & 3.8 &  & \\
         & $K_{1H}\bar{K}$ & $P$ & x &  & \\
         & $K(1460)\bar{K}$ & $S$ & \o & \\
$\Gamma_{tot}$ & & & \textbf{80.5} &  & \\
\hline
&  &  & ~~$H_2~(2.015~$GeV)~~  & $~~H_4~(2.225~$GeV)~~  \\
\hline
$~2^{+-}~$ & $\omega\pi$ & $D$ &  x & x & \\
         & $\rho\eta$ & $D$ &  0.3 & 0.15 \\
         & $\rho\eta'$ & $D$ & 0.1 & x \\
         & $b_1(1235)\eta$ & $P$ & 1.5 & 8.9 &  \\
         &  & $F$ & x & 0.2 & \\
         & $h_1(1170)\pi$ & $P$ & 5.5 & 18.4 &\\
         &  & $F$ & 0.1 & 1.0 & \\
         & $a_1(1260)\pi$ & $P$ & 8.0 & 9.4 &  \\
         &  & $F$ & x & 0.1 & \\
         & $a_2(1320)\pi$ & $P$ & 6.5 & 0.2\\
         &  & $F$ & 0.1 & 0.1 & \\
         & $f_1(1285)\rho$ & $P$ & \o & 15.3 \\
         &  & $F$ & \o & x & \\
         & $f_2(1270)\rho$ & $P$ & \o & 10.2 \\
         &  & $F$ & \o & x & \\
         & $\omega(1420)\pi$ & $D$ & x & x \\
         & $\pi(1300)\pi$ & $D$ & x & 0.15 \\
         & $K^*\bar{K}$ & $D$ & x & x \\
         & $K_{1L}\bar{K}$ & $P$ & x & x & \\
         &  & $F$ & x & x &\\
         & $K_{1H}\bar{K}$ & $P$ & 0.2 & 2.6 & \\
         &  & $F$ & x & x &\\
         & $K^*(1410)\bar{K}$ & $D$ & x & x\\
         & $K(1460)\bar{K}$ & $D$ & x & x\\
         & $K_2^*(1430)\bar{K}$ & $P$ & 0.5 & x \\
         &  & $F$ & x & x & \\
$\Gamma_{tot}$ & & & \textbf{22.8} & \textbf{66.9} & \\ 
\hline\hline
\end{longtable}

\begin{figure}
\includegraphics[width=0.80\linewidth]{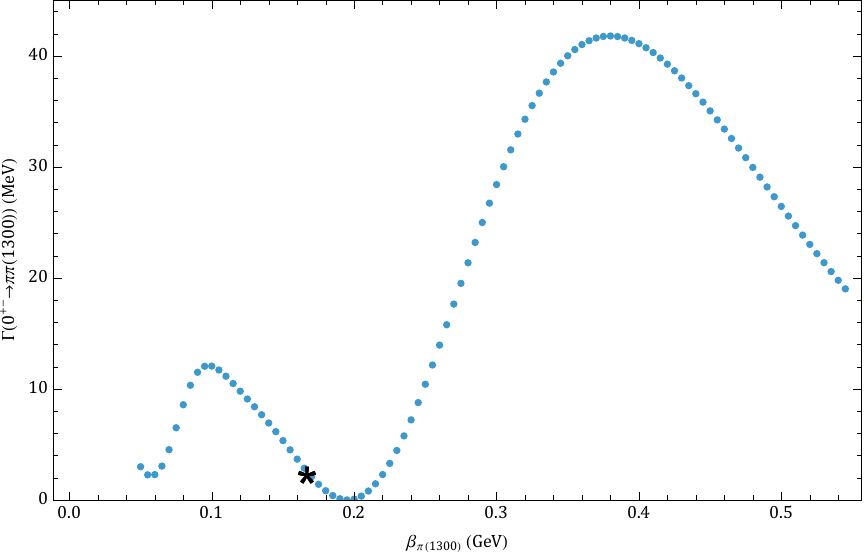}
\caption{Plot of $\Gamma(0^{+-}\rightarrow\pi(1300)\pi)$ as a function of the $\pi(1300)$ $\beta$ parameter. The asterisk corresponds to the value of $\Gamma$ at $\beta_{\pi(1300)}=0.167$, which is what we obtain from the spin dependent model in Eq. \ref{eq:SDmodel} and which is reported in Table \ref{tab:IsovectorDecays}.}
\label{fig:pi1300}
\end{figure}

The results show that the three isovectors are all narrow. While the $2^{+-}$ states were previously reported to be narrow, the $0^{+-}$ comes as a surprise. If we compare our results to those of previous analyses such as the  PPS model \cite{Page:1998gz} (and the IKP model reported therein) or the work of Ref. \cite{Chen2025} which uses a very similar approach to ours, the $0^{+-}$ was always reported as being wide, with a total width in the 250-650 MeV range and a dominant contribution of about 200-250 MeV coming from the $\pi(1300) \pi$, which in our work is almost entirely suppressed. In our model, the suppression of the channel is strongly dependent on the value of the $\beta_m$ parameter used for the $\pi(1300)$. This dependence is shown in Figure \ref{fig:pi1300}, where the total width is plotted as a function of $\beta_{\pi(1300)}$. Oscillatory behavior like the one in the figure are typical of decay channels that include excited states.

Fitting the light ordinary meson spectrum (via Eq. \ref{eq:SDmodel}) gives $\beta_{\pi}=0.54$ Gev and $\beta_{\pi(1300)}=0.167$ GeV. Indeed, all excited states have relatively small $\beta$ parameters since $\beta$ is inversely proportional to the RMS radius of of meson. Remarkably, the preferred value of $\beta_{\pi(1300)}=0.167$ GeV lies very close to a node in the decay amplitude, which occurs at about $\beta_{\pi(1300)}=0.195$ GeV. The figure also shows that the decay width rapidly increases with increasing $\beta$, reaching its maximum of about $42$ MeV at $\beta_{\pi(1300)}=0.38$ GeV. Thus, even our maximal partial width is  smaller than those of \cite{Page:1998gz,Chen2025} by a factor of at least six. 
It is perhaps useful to point out that, in contrast to our method, authors often employ a single value for the $\beta$ parameters of ground state and excited mesons; as we see here, such an assumption is ill advised.

For all three states, $S+P$ wave decays are preferred. This is the case for the $0^{+-}$, whose main decay channels are the $h_1(1170) \pi$ and $b_1(1235) \eta$, in that order. The same is also true for the $2^{+-}$ and $2^{+-'}$ states. However, these states show some differences. The $2^{+-}$ decays mainly into $a_1(1260)\pi$, $a_2(1320)\pi$ and $h_1(1170) \pi$, in that order, but has a small $b_1(1235) \eta$ contribution. In contrast, the main decay channels for the $2^{+-'}$ are $h_1(1170) \pi$ and $a_1(1260)\pi$. In this case, $b_1(1235) \eta$ is more significant, but the $a_2(1320)\pi$ is negligible. 

These differences, we suspect, are not only due to the difference in hybrid mass between the two states, but rather due to the different value of total hybrid angular momentum, $L$, ($L=1$ for the $H_2$ and $L=2$ for the $H_4$) which in turn leads to different expressions for the decay amplitudes. Interestingly, the $2^{+-'}$ also receives large contributions from the $f_1\rho$ and $f_2\rho$ channels, $P+P$ modes, which are below threshold for the lighter $2^{+-}$. Overall, the total widths for the $2^{+-}$ states lie somewhere in the middle of the values reported in \cite{Page:1998gz} and \cite{Chen2025}.

The partial widths obtained in the model generally follow the pattern of $^3S_1$ models, in which the leading partial wave is much larger than sub-leading ones, as seen for all the channels that allow decays into $P$ and $F$ waves.

\subsection{Isoscalars}
\label{sect:isoscalars}

We consider isoscalar hybrids to be linear combinations of light and strange quarks defined by
\begin{eqnarray}
    \ket{L}&=&\cos{\theta_{I}}\ket{n\bar{n}}-\sin{\theta_{I}}\ket{s\bar{s}} \nonumber \\
    \ket{H}&=&\sin{\theta_{I}}\ket{n\bar{n}}+\cos{\theta_{I}}\ket{s\bar{s}},
\end{eqnarray}
where $\theta_{I}$ is the mixing angle. Currently, there is no experimental information about these mixing angles. The lattice computation cited above does, however, report light hybrid mixing angles\cite{Dudek:2013yja}. The computation shows that there is very little light/strange mixing outside of the lightest $H_1$ multiplet. This agrees  with the results presented in \cite{Swanson:2023zlm} where the excited states are expected to be ideally mixed. Hence, we set $\theta_I=0^\circ$ for all of them. This implies that heavy ($\ket{s\bar{s}g}$) isoscalars can only decay to pairs of mesons with a strange-quark component, such as kaon-pairs, or $\phi\eta$ final states. Table \ref{tab:IsoscalarDecays} reports the decay widths of the six isoscalars considered here. 

\begin{center}
\begin{longtable}{c|c|c|ccc}
\caption{Strong decays of the six isoscalar states. The number in parenthesis is the hybrid mass. The meaning of the symbols is as follows: x = negligible, $\o=$ threshold, - = forbidden.}
\label{tab:IsoscalarDecays}\\
\hline\hline
Hybrid $J^{PC}$ & ~Decay Channel~ & ~~Wave~~ & $~~~~~~~\ket{n\bar{n}g}~~~~~~~$  & $~~~~~~~\ket{s\bar{s}g}~~~~~~~$ &  \\
 & & & ~(1.955 GeV)~ & ~(2.055 GeV)~ & \\
\hline
$~0^{+-}~(H_2)$ & $b_1(1235)\pi$ & $P$ &  171.3 & - & \\
               & $h_1(1170)\eta$ & $P$ & 20.6  & - & \\
               & $K_{1L}\bar{K}$ & $P$ & 12.5  & 27.3 & \\
               & $K_{1H}\bar{K}$ & $P$ &  0.2 & 1.8 & \\
               & $K(1460)\bar{K}$ & $S$ &  \o & 4.3 & \\
$\Gamma_{tot}$ & & & \textbf{204.6} & \textbf{33.4} & \\ 
\hline
 & & & ~(2.035 GeV)~ & ~(2.125 GeV)~ & \\
\hline
$~2^{+-}~(H_2)$ & $\omega\eta$ & $D$ &  x & - & \\
         & $\rho\pi$ & $D$ &  1.3 & - \\
         & $\omega\eta'$ & $D$ & x & - \\
         & $\phi\eta$ & $D$ & - & x \\
         & $\phi\eta'$ & $D$ & - & x \\
         & $b_1(1235)\pi$ & $P$ & 16.0 & - &  \\
         &  & $F$ & 0.4 & - & \\
         & $h_1(1170)\eta$ & $P$ & 2.2 & - &\\
         &  & $F$ & x & - & \\
         & $\omega(1420)\eta$ & $D$ & x & - &  \\
         & $\rho(1450)\pi$ & $D$ & x & - \\
         & $a_1(1260)\rho$ & $P$ & 4.3 & - \\
         &  & $F$ & x & - & \\
         & $a_2(1320)\rho$ & $P$ & \o & - \\
         &  & $F$ & x & - & \\
         & $K^*\bar{K}$ & $D$ & x & 0.5 \\
         & $K_{1L}\bar{K}$ & $P$ & 0.9 & 4.0 & \\
         &  & $F$ & x & x & \\
         & $K_{1H}\bar{K}$ & $P$ & 0.9 & 6.9 & \\
         &  & $F$ & x & x & \\
         & $K^*(1410)\bar{K}$ & $D$ & x & x \\
         & $K_2^*(1430)\bar{K}$ & $P$ & 0.8 & 8.1 \\
         &  & $F$ & x & x & \\
$\Gamma_{tot}$ & & & \textbf{26.8} & \textbf{19.5} & \\  
\hline
 & & & ~(2.255 GeV)~ & ~(2.335 GeV)~ & \\
\hline
$~2^{+-}~(H_4)$ & $\omega\eta$ & $D$ & x & - & \\
         & $\rho\pi$ & $D$ &  0.5 & - \\
         & $\omega\eta'$ & $D$ & x & - \\
         & $\phi\eta$ & $D$ & - & x \\
         & $\phi\eta'$ & $D$ & - & x \\
         & $b_1(1235)\pi$ & $P$ & 55.8 & - &  \\
         &  & $F$ & 2.7 & - & \\
         & $h_1(1170)\eta$ & $P$ & 9.7 & - &\\
         &  & $F$ & 0.3 & - & \\
         & $\rho(1450)\pi$ & $D$ & x & - \\
         & $a_1(1260)\rho$ & $P$ & 66.9 & - \\
         &  & $F$ & 0.2 & - & \\
         & $a_2(1320)\rho$ & $P$ & 29.7 & - \\
         &  & $F$ & x & - & \\
         & $K^*\bar{K}$ & $D$ & x & 0.1 \\
         & $K_{1L}\bar{K}$ & $P$ & x & 0.1 & \\
         &  & $F$ & x & x & \\
         & $K_{1H}\bar{K}$ & $P$ & 2.9 & 7.4 & \\
         &  & $F$ & x & x & \\
         & $K^*(1410)\bar{K}$ & $D$ & x & x \\
         & $K_2^*(1430)\bar{K}$ & $P$ & 0.1 & 0.5 \\
         &  & $F$ & x & x & \\
         \hline
$\Gamma_{tot}$ & & & \textbf{168.8} & \textbf{8.1} & \\ 
\hline\hline
\end{longtable}
\end{center}

The light $0^{+-}$ isoscalar hybrid appears to be broad, with a very large contribution from the $b_1(1235) \pi$ channel and smaller, but significant, contributions from $h_1(1170)\eta$ and $K_{1L}\bar{K}$, and it might therefore be hard to detect. On the other hand, the strange $0^{+-}$ is narrow, and could appear as a sharp peak in the $K_{1L}\bar{K}$ channel. The $K(1460)\bar{K}$ channel is below threshold for the light state and negligible for the strange state. Again, the reason for this lies in the small value of the $\beta_m$ parameter obtained for the $2^1S_0$ $K(1460)$ state, just as it was the case for the $\pi(1300)$ above, which appears to be something unique to our model. In fact, both \cite{Page:1998gz} and \cite{Chen2025} report very large contributions coming from $K(1460)\bar{K}$, with the latter reporting a branching ratio of $\frac{\Gamma(K(1460)\bar{K})}{\Gamma_{tot}}=\frac{394.4}{484}=81.5\%$.

The light $2^{+-}$ is also narrow, with its largest contribution coming from the $b_1(1235)\pi$ channel, plus a small $a_1(1260)\rho$ component, if its mass lies below the $a_2(1320)\rho$ threshold. If, however, the hybrid mass is pushed above that threshold, the $a_2(1320)\rho$ partial width becomes an important contribution to the total width. The strange $2^{+-}$ is even narrower, and the only significant channels are the $K_{2}^{*}(1430)\bar{K}$, $K_{1H}\bar{K}$, and $K_{1L}\bar{K}$, in that order.

The light $2^{+-'}$ is broader, and has somewhat idiosyncratic decay properties. While all the states seem to obey the $S+P$ rule mentioned above, according to which $S+P$ decay modes are dominant, the light $2^{+-'}$ main decay channel is $a_1(1260)\rho$, which is a $P+P$ mode. This is followed by $b_1(1235)\pi$, an $S+P$ mode, and $a_2(1320)\rho$, another $P+P$ channel. Perhaps, these distinctive decay properties could help in its experimental detection.

The strange $2^{+-'}$ is the narrowest state, with $K_{1H}\bar{K}$ being the only significant decay mode. Our predictions here are more in line with the PSS model, while Ref. \cite{Chen2025} reports larger widths.

We remark that the production of all six of the states considered here will be suppressed at COMPASS because the high pion beam energy implies that pomeron exchange dominates production. This selects isobars with $I^G = I^G(\pi) =1^-$. Thus the $\pi_1$ can be produced, but  $b_0$ [$I^G(J^{PC}) = 1^+(0^{+-})$], $h_0$ [$I^G(J^{PC}) = 0^-(0^{+-})$], $b_2$ [$I^G(J^{PC}) = 1^+(0^{+-})$], and $h_2$ [$I^G(J^{PC}) = 0^-(2^{+-})$] are all disfavored. In contrast, GlueX, with its $I^G= 0^- + 1^+$ beam, can produce both the isovector and isoscalar hybrids with exchanges on the pomeron and other trajectories. Thus GlueX may be able to explore the entire set of low lying exotic hybrids.

\section{Summary and Conclusions}
\label{sect:conc}

We have developed a model of hybrid structure and light hybrid decays that leverages the QCD Hamiltonian in Coulomb gauge and models the relevant gluodynamics in terms of a dynamically generated constituent gluon. This approach finds support in lattice field computations, which determine that an axial gluonic degree of freedom is sufficient to describe the main features  of the low lying hybrid spectrum. The results presented here are compared to previous work, such as the flux-tube PSS model \cite{Page:1998gz} and a recent constituent gluon model by Chen and Liu \cite{Chen2025} which is in many regards very similar to our approach. We find that our results mostly agree with those papers, with some exceptions. The most notable difference is the $0^{+-}$ which is very narrow in our computation, due to the suppression of the decay mode $0^{+-}\rightarrow 2^1S_0+1^1S_0$. The works cited above find that particular decay mode to make the largest contribution to the total width, both for the isovector and the strange isoscalar.

We remark that the results obtained here are dependent on many parameters that come with uncertainty. First of all, the small value of the $2^1S_0+1^1S_0$ in the $0^{+-}$ ($\pi(1300)\pi$ for the isovector and $K(1460)\bar{K}$ in the strange isoscalar)  is dependent on the wavefunction parameter $\beta_m$ for the $2^1S_0$, which in turn is determined by the spin-dependent model used to fit the meson masses.

Additionally, the total widths are clearly dependent on the hybrid masses used. The masses were crudely extrapolated from the lattice computations by shifting all of them down by the same amount so that the $1^{-+}$ mass agrees with the mass of the $\pi(1600)$, which was set at  1600 MeV. Choosing different masses would clearly change the total widths  and could allow additional channels to go above or below threshold. This is a very important point in assessing the reliability of the results, because, as obtained by lattice computations, all the states of interest are actually decaying resonances and the masses determined should all be considered to have a systematic error at least as big as the (unknown) hadronic width of the states \cite{jjd}.

In order to do a sensitivity analysis of our results, we computed the width variation as a function of several of the parameters that are used in the model for the isovector $0^{+-}\rightarrow\pi\pi(1300)$,  the $H_2$ isoscalar $2^{+-}\rightarrow b_{1}(1235)\pi$,  and for the total width of the $H_4$ isovector $2^{+-}$. The parameters varied were the Coulombic strength $\alpha_S$, the hybrid mass $M_{hyb}$, the hybrid variational parameters $\alpha_{hyb}$ and $\beta_{hyb}$, and the meson parameters $\beta_{m}$. In each case, we varied each parameter by $\pm10\%$, but for the total  $2^{+-}$ width, all the meson betas were set to be equal, ranging from $0.30$ to $0.50$ GeV. The results are shown in Figures \ref{fig:pipiVar}, \ref{fig:pib1Var}, and \ref{fig:TwoPMVar}. As the plots show, the decays are most sensitive to the $\beta$ parameters and the hybrid masses. The dependence on $\alpha_S$ is simple -- $\Gamma$ is largely proportional to $\alpha_S$ -- but less significant. These results are expected in large part. A higher hybrid mass increases the width value by increasing the available phase space for the decay and also by allowing additional channels to open up. A variation in the beta parameters for both the hybrids and conventional mesons, affects the widths by changing the value of the Gaussian functions in the radial expansion, which depending on the $\beta$ value, can have a large effect on the width.

\begin{center}
\begin{figure}[ht]
\includegraphics[width=0.80\linewidth]{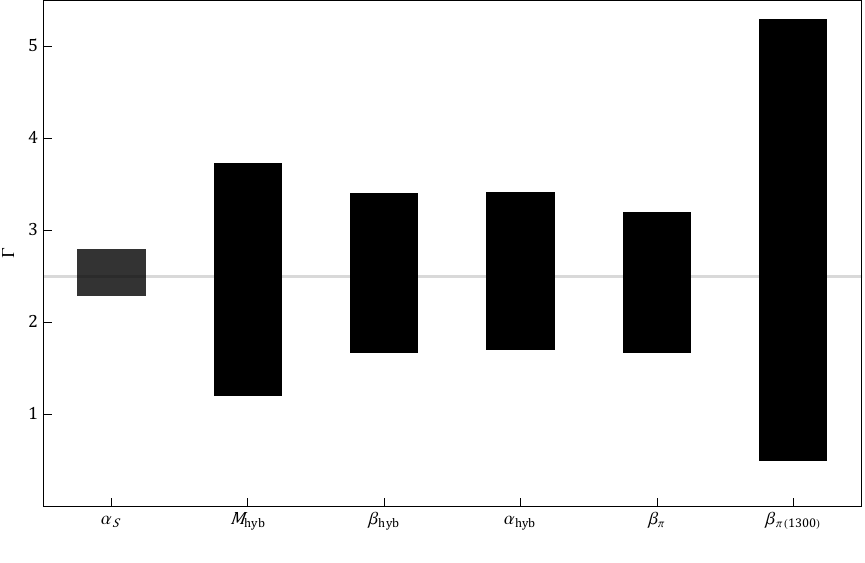}
\caption{Width variation of $0^{+-}\rightarrow\pi\pi(1300)$ as a function of model parameters. Parameters were varied by 10\%.}
\label{fig:pipiVar}
\end{figure}

\begin{figure}[ht]
\includegraphics[width=0.80\linewidth]{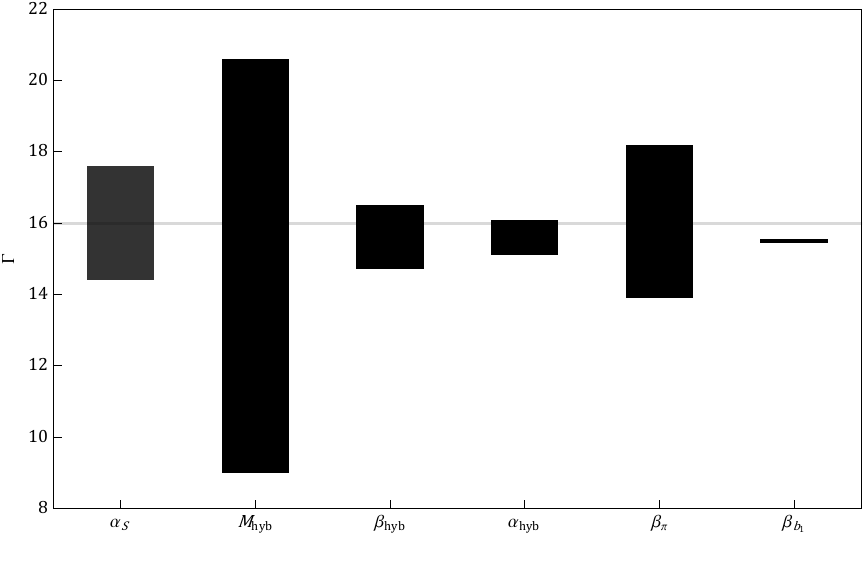}
\caption{Width variation of the $H_2$ isoscalar $2^{+-}\rightarrow b_{1}(1235)\pi$ as a function of model parameters. Parameters were varied by 10\%.}
\label{fig:pib1Var}
\end{figure}

\begin{figure}[ht]
\includegraphics[width=0.80\linewidth]{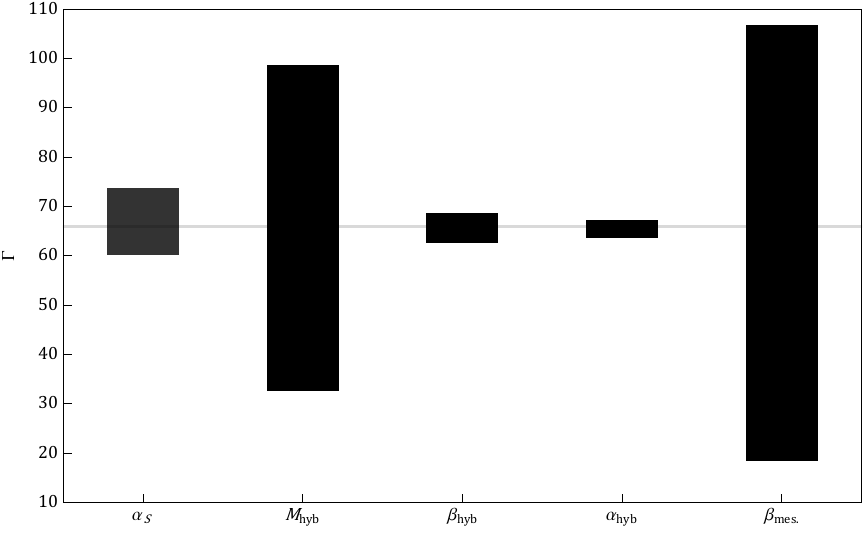}
\caption{Total width variation of the $H_4$ isovector $2^{+-}$ as a function of model parameters. Parameters were varied by 10\%, except for the $\beta_{mes.}$ which were varied between 0.30 GeV and 0.50 GeV.}
\label{fig:TwoPMVar}
\end{figure}
\end{center}

We note that all of the strange isoscalar hybrids appear very narrow. This is in part because we have chosen ideal light-strange mixing, which only allows kaon-pair decay modes.

To our knowledge, there is currently no experimental evidence of $0^{+-}$ exotic states, but preliminary results for the reaction $\pi^{-}p\rightarrow2\pi^{-}2\pi^{+}\pi^{0}$ by the BNL(E852) collaboration, found a strong signal for a $I^{G}J^{PC}=0^{-}2^{+-}$ state near 1.9 GeV (labeled $h_2$) in the process $\pi^{-}p \rightarrow h_2n\rightarrow b_1\pi n\rightarrow 2\pi\omega \rightarrow 5\pi n$ \cite{Adams:2005tx}. This could be a candidate for our isovector $2^{+-}$ state, whose mass is very close to 1.9 GeV and whose main decay mode is the $b_1(1235)\pi$ channel. However, these preliminary results need  confirmation and should therefore be taken with caution. Searches at GlueX 
and in multi-pion final states in $J/\psi$ and $\chi_{c1}$ decay at BES appear to be promising for providing such confirmation.

The most obvious continuation of the present work would be to compute the strong decays of the remaining states in Table \ref{tab:JPC}. These hybrid states all have conventional quantum numbers ($0^{++}$,$2^{++}$,$1^{+-}$, $3^{+-}$) and can therefore mix with conventional mesons, making their identification more challenging. Hence, computing their strong decays might reveal some interesting features that could distinguish them from conventional mesons and thus aid in their detection. Additionally, the model presented here can be extended to include spin-dependent terms, which in turn could be used to compute the mass spectrum of hybrids.

\acknowledgments

Swanson
acknowledges support by the U.S. Department of Energy under contract DE-SC0019232.
This work contributes to the goals of the US DOE ExoHad Topical Collaboration, Contract
No. DE-SC0023598.

\end{document}